\documentclass[11pt]{article}
\def\be{\begin{equation}}
\def\ee{\end{equation}}
\def\ba{\begin{eqnarray}}
\def\ea{\end{eqnarray}}
\def\nn{\nonumber}
\def\lb{\label}
\def\bb{\bibitem}
\def\ol{\overline}
\def\E{{\cal E}}
\def\2{\sqrt2}

\begin{document}

\begin{titlepage}

\date{}

\title{
\begin{flushright}\begin{small}    LAPTH-047/26
\end{small} \end{flushright} \vspace{1cm}
Comment on ``Multi-black holes in Bertotti-Robinson spacetime''}

\author{G\'erard Cl\'ement\thanks{Email: gclement@lapth.cnrs.fr} \\ \\
{\small LAPTh, Universit\'e Savoie Mont Blanc, CNRS, F-74940  Annecy, France}}

\maketitle

\begin{abstract}
We point out that the asymptotically Bertotti-Robinson multi-black hole solutions of Einstein-Maxwell theory can be constructed using only three-dimensional sigma-model techniques, without resorting to the monodromy-matrix approach.
\end{abstract}
\end{titlepage}
\setcounter{page}{2}

In \cite{furugori}, exact stationary solutions of the Einstein-Maxwell equations describing multi-black holes in the Bertotti-Robinson spacetime were recently constructed using the monodromy-matrix formalism. This powerful formalism is associated with the two-dimensional integrable sigma model valid for the stationary axisymmetric sector of Einstein-Maxwell theory. We wish to point out that the same solutions can be directly generated by using only three-dimensional sigma-model techniques first developed forty years ago \cite{spat}, without resorting to the cumbersome monodromy-matrix approach.

We first briefly summarize the Ernst \cite{ernst} approach to the sigma-model formulation of the stationary sector of Einstein-Maxwell theory. The equations derive from the action
\begin{equation}
S=\frac{1}{16\pi}\int[-R - F_{\mu\nu}F^{\mu\nu}] \sqrt{-g}\,d^4x,
\end{equation}
where $F = dA$. Assuming the existence of a timelike Killing vector $\partial_t$, the reduction to three dimensions is achieved by the ansatz
\be\lb{an}
ds^2 = f(dt - \omega_i dx^i)^2 - f^{-1}h_{ij}dx^idx^j,
 \ee
where the fields $f$, $\omega_i$ and $h_{ij}$ depend on the space coordinates $x^i$ ($i=1,2,3$). The electric potential is $v=A_0$, and the scalar magnetic $u$ and twist $\chi$ potentials are defined by the duality relations
\ba
\partial_i u &=& \frac1{2f}\,h^{-1/2}h_{ij}\,\epsilon^{jkl}F_{kl}, \nn\\
\partial_i\chi &=& -f^2\,h^{-1/2}h_{ij}\,\epsilon^{jkl}\partial_k\omega_l + 2(u\partial_i v - v\partial_i u).
\ea
The complex Ernst potentials are related to the four real potentials $f$, $\chi$, $v$ and $u$ by
\be
{\cal E} = f + i \chi - \ol{\psi}\psi\,, \qquad \psi = v + iu\,.
\ee
Then one can show \cite{mg} that the stationary Einstein-Maxwell equations derive from the three-dimensional SU(2,1)/S(U(2)
$\times$ U(1)) sigma-model action
\be
S = \frac12\int\left[R^h + \frac14{\rm Tr}(\nabla M \nabla M^{-1})\right]\sqrt{h}\,d^3x,
\ee
where $R^h$ is the three-dimensional Ricci scalar, and $M$ is the coset matrix representative
\begin{equation} \label{MG}
M = f^{-1} \pmatrix{
1 & \sqrt{2}\,\psi & i(\overline{\cal E} - {\cal E} + 2\psi\overline{\psi})/2 \cr
\sqrt{2}\,\overline{\psi} & -({\cal E} + \overline{\cal E} - 2\psi\overline{\psi})/2 &
-i\sqrt{2}\, {\cal E}\overline{\psi} \cr
i(\overline{\cal E} - {\cal E} - 2\psi\overline{\psi})/2 & i\sqrt{2}\, \overline{\cal E}\psi &
{\cal E}\overline{\cal E} }.
\end{equation}
This matrix is hermitean, $M^+ = M$, and belongs to SU(2,1), $M^+JM = J$,
${\rm det}M = 1$, with
\begin{equation}
J = \pmatrix{
0 & 0 & -i \cr
0 & 1 & 0 \cr
i & 0 & 0 }.
\end{equation}

From a given solution $M(x)$, one can generate another solution $M_P(x)$ by the action of an SU(2,1) matrix transformation $P$,
\be
M_P = P^+MP.
\ee
This transfotmation leaves invariant the three-dimensional metric $h_{ij}$.

The transformations $P$ belonging to the isotropy subgroup S(U(2)$\times$ U(1)) (including the well-known Ehlers and Harrison transformation) leave invariant the asymptotic matrix $M(\infty) = \eta$. For asymptotically locally flat solutions (the simplest of which is the Schwarzschild solution), this asymptotic matrix is
\be
\eta_S = \pmatrix{
1 & 0 & 0 \cr
0 & -1 & 0 \cr
0 & 0 & 1 }\,.
\end{equation}
Other group transformations than those belonging to those of the isotropy subgroup can be used to generate non-asymptotically flat solutions from an asymptotically flat seed. As observed in \cite{kerr}, the Schwarzschild metric and the open Bertotti-Robinson metric share the same reduced spatial metric $h_{ij}$. There is therefore an SU(2,1) transformation generating the Bertotti-Robinson solution from the Schwarzschild
seed. In terms of the Ernst potentials, this is the involution $P_{BS}:(\E,\psi)\; \leftrightarrow \; (\hat\E,\hat\psi)$:
\be\lb{invo}
\hat{\cal E} = \frac{-1 + {\cal E} + 2 \psi}{1 - {\cal E} + 2 \psi}\,, \qquad
\hat{\psi} = \frac{1 + {\cal E}}{1 - {\cal E} + 2 \psi}\,,
\ee
The matrix implementation of this transformation
\be
P_{BS} = P_{SB} = -\frac12\pmatrix{
1 & \2 & i \cr
\2 & 0 & -i\2 \cr
-i & i\2 & 1 }
\ee
leads from $\eta_S$ to the asymptotic matrix $\hat{M}(\infty) = \eta_B$,
\be\lb{etab}
\eta_B = \pmatrix{
0 & 0 & i \cr
0 & 1 & 0 \cr
-i & 0 & 0 }\,,
\end{equation}
of the Bertotti-Robinson solution.

In the case of solutions depending on a single real potential $\sigma(\vec{x})$, it follows from the equations of motion that this potential can be chosen to be harmonic,
\be\lb{harm}
\nabla^2\sigma = 0.
\ee
The solutions are then target space geodesics, which may be written, under the assumption $\sigma(\infty)=0$,
\be\lb{geo}
M(\vec{x}) = \eta \,{\rm e}^{A\sigma(\vec{x})},
\ee
with $A \in su(2,1)$, implying Tr$(A) = 0$. The three-dimensional Einstein equations then reduce to
\be
R^h_{ij} = \frac14{\rm Tr}(A^2)\partial_i\sigma\partial_j\sigma.
\ee
If furthermore
\be
{\rm Tr}(A^2) = 0
\ee
(implying also $A^3=0$ \cite{bps})), the reduced three-space is flat. The harmonic condition (\ref{harm}) then reduces to the linear Laplace equation, which admits among others the multi-monopole solutions
\be\lb{multi}
\sigma(\vec{x}) = \sum_\alpha\frac{c_\alpha}{|\vec{x}-\vec{a_\alpha}|}.
\ee

In the case of asymptotically flat solutions ($\eta = \eta_S$), this construction \cite{bps} is equivalent to the Majumdar-Papapetrou \cite{mp} construction of multi-extreme black holes. In the special case of electric solutions (the generalization to dyonic solutions is straightforward), the charge matrix $A$ for extreme Reissner-Nordstr\"om black holes is\footnote{rescaled by a factor $1/2$ from (B11) of \cite{bps}, with $\alpha=\beta=0$.}
\be
A = \frac12\pmatrix{
2 & \2 & 0 \cr
-\2 & 0 & i\2 \cr
0 & i\2 & -2 }\,.
\end{equation}
Its lift to the Bertotti-Robinson domain is
\be
\hat{A}  = \pmatrix{
0 & 0 & 0 \cr
\2 & 0 & 0 \cr
0 & -i\2 & 0 }\,,\quad {\hat{A}}^2 = 2\pmatrix{
0 & 0 & 0 \cr
0 & 0 & 0 \cr
-i& 0 & 0 }\,.
\end{equation}
The resulting multi-center asymptotically Bertotti-Robinson coset matrix is
\be
\hat{M} = \eta_B {\rm e}^{\hat{A}\sigma} = \pmatrix{
\sigma^2 & \2\sigma & i \cr
\2\sigma & 1 & 0 \cr
-i& 0 & 0 }\,,
\end{equation}
corresponding to the Ernst potentials
\be
\hat\E = 0, \quad \hat\psi = \sigma^{-1},
\ee
and to the multi-center metric
\be\lb{multibr}
d\hat{s}^2 = \sigma^{-2}dt^2 - \sigma^{2}d{\vec{x}}^2,
\ee
$\sigma(\vec{x})$ being given by (\ref{multi}). Equation (\ref{multibr}) reproduces (88) of \cite{furugori}, without the assumption of axial symmetry.

Let us stress that, in the three-dimensional sigma-model approach, this multi-Bertotti Robinson solution may be obtained in two ways. Either horizontally, by constructing the coset matrix $\hat{M} = \eta_B {\rm e}^{\hat{A}\sigma}$, with $\eta_B$ given by (\ref{etab}) and $\hat{A}$ constrained by ${\hat{A}}^+J + J{\hat{A}} = 0$, ${\hat{A}}^+\eta_B - \eta_B{\hat{A}}= 0$ and ${\rm Tr}{\hat{A}} = {\rm Tr}{\hat{A}}^2 = 0$. Or vertically, by the lift $\hat{M} = P_{BS}^+MP_{BS}$ to the Bertotti-Robinson domain of the Majumdar-Papapetrou solution\footnote{An even more direct way, related to the near-horizon limit, is to carry out on the Majumdar-Papapetrou solution the translation $\sigma\to\sigma-1$, together with a gauge transformation.}
\be
ds^2 = -(1+\sigma)^{-2}dt^2 + (1+\sigma)^2d{\vec{x}}^2,  \quad A = \frac\sigma{1+\sigma}dt.
\ee

Let us also comment on the form of equation (90) of \cite{furugori}),
\be
\sigma(\vec{x}) = \sum_{j=1}^N\frac{m_j}{|\vec{x}-\vec{x_j}|} + \sum_{j=1}^M\frac{1/B_j}{|\vec{x}-\vec{x^B_j}|}.
\ee
While the first sum is over extremal black holes, ``for the second family, the positions are determined by the parameters $B_j$, which characterize the Bertotti-Robinson background'' (beginning of subsection V.A of \cite{furugori}). Actually, all the terms in the sum (\ref{multi}) refer to extreme black holes. The asymptotic Bertotti-Robinson behaviour is not linked with any particular term in this sum (all the terms in $B_j$ could be discarded by taking the limit $B_j \to\infty$), but derives from the form of the asymptotic coset matrix (\ref{etab}).

Finally, the generalization to Israel-Wison-Perj\`es \cite{iwp} solutions with Bertotti-Robinson asymptotics is carried out in section VI of \cite{furugori}. Again, this can be obtained directly by the Bertotti-Robinson lift of the three-dimensional coset matrix given in (B16) of \cite{bps}, leading to
\begin{equation} \label{iwp}
\hat{M} = \eta_B\,{\rm e}^{\hat{A}\sigma + \hat{B}\tau} = \eta_B[1 + \hat{A}\sigma + \hat{B}\tau + \frac{1}{2}{\hat{A}}^2(\sigma^2 + \tau^2)]\,,
\end{equation}
where $\sigma$ and $\tau$ are arbitrary harmonic potentials, and $\hat{A}$ and $\hat{B}$ are the lifts of the charge matrices rescaled by a factor $2$ from those given in (B11) and (B14) of \cite{bps}. The corresponding Ernst potentials, lifted from (B18) of \cite{bps} (rescaled by a factor $2$) by the Bertotti-Robinson lift (\ref{invo}), is
\be
\hat{\E} = 4\frac{1-{\rm}e^{i\gamma}}{1+{\rm}e^{i\gamma}}, \quad \hat{\psi} = 2\frac{e^{i\beta}}{(1+{\rm}e^{i\gamma})\zeta},
\ee
where $\zeta=\sigma-i\tau$ is a complex harmonic function, and $\gamma=\alpha+\beta$. The resulting four-dimensional solution is (117) of \cite{furugori}.

\section*{Acknowledgments}

I thank D. Gal'tsov for a critical reading of the manuscript. I also thank the authors of \cite{furugori} for an informative discussion.

\end{document}